\documentclass[11pt]{article}

\usepackage{amsmath,amssymb,amsthm}
\usepackage{float}
\usepackage{latexsym,amssymb, amsthm, epsfig}

\def\qsd{{\sc{qsd }}}

\begin{document}
\noindent

\def\E{{\mathbb E}} 
\def\P{{\mathbb P}} 
\def\R{{\mathbb R}} 
\def\Z{{\mathbb Z}} 
\def\V{{\mathbb V}} 
\def\N{{\mathbb N}} 
\def\cL{{\mathcal{L}}}  
\def\one{{\mathbf 1}}  
\def\ind{{\mathbf 1}}
\def\qsd{{\sc{qsd\,}}}  
\def\qsds{{{\sc{qsd}}s\,}}
\def\fv{{\sc{fv\,}}}  
\def\reff#1{(\ref{#1})}
\def\square{\ifmmode\sqr\else{$\sqr$}\fi}
\def\sqr{\vcenter{
         \hrule height.1mm
         \hbox{\vrule width.1mm height2.2mm\kern2.18mm\vrule width.1mm}
         \hrule height.1mm}}                  % This is a slimmer sqr.
         
\def\rhohatt{{\hat\rho^N_t}}

\def\rhohattc{{\hat\rho^{N,c}_t}}

\def\Ehatt {{\frac{\E\hat\eta^{N,\mu}_t}{N} }}    

\def\rhohat{{\hat\rho^N}}
\def\rhohat{{\hat\rho^N}}

\def\Ehat{{ \frac{\E\hat\eta^{N}}{N} }}

 \theoremstyle{plain}
\newtheorem{lema}{Lemma}[section]
\newtheorem{teo}{Theorem}[section]
\newtheorem{propo}{Proposition}[section]
\newtheorem{alg}{Algorithm}
\newtheorem{coro}{Corolary}[section]
 \newtheorem{conj}{Conjecture}

\theoremstyle{definition}
\newtheorem{de}{Definition}
\newtheorem{ex}{Example}
\newtheorem{al}{Algorithm}

\theoremstyle{remark}
\newtheorem{obs}{Remark}[section]

\begin{center}
{\Large\bf Fleming-Viot particle system driven by a random walk on $\mathbb{N}$ }
\end{center}

\vspace{0.5cm}

\centerline{\bf Nevena Mari\'{c} } 

\centerline{\emph{University of Missouri - St. Louis}}
 
\vspace{0.5cm}

%\pagenumbering{roman} 
%\thispagestyle{plain} 
%\begin{center}  
\noindent {\large\bf Abstract}  
%\end{center}  
Random walk on $\N$ with negative drift and absorption at 0, when conditioned on survival, has uncountably many invariant measures (quasi-stationary distributions, \qsd) $\nu_c$. We study a Fleming-Viot(\fv) particle system driven by this process and show that  mean normalized densities of the {\fv} unique stationary measure converge to the minimal {\qsd,} $\nu_0$, as $N \to \infty$.  Furthermore, every other \qsd of the random walk   ($\nu_c$, $c>0$) corresponds to a metastable state of the \fv particle system.
\paragraph{Keywords} Quasi-stationary distributions. Fleming-Viot process. Selection principle. Metastability.

%\paragraph{AMS Classification} 60F 60K35

\section{Introduction}
\label{intro}
In the Fleming-Viot (\fv) particle system there are $N$ ($N >1$) particles where each particle evolves as a Markov chain  $Z_t$ which we call the {\em driving process}. The assumption is that $Z_t$ is irreducible on a countable state space and has an absorbing state.  As soon as one particle is absorbed, it reappears immediately, choosing a new position according to the empirical measure at that time. Between the absorptions, the particles move independently of each other.  Our focus is on the relation of empirical measures of the \fv process with  quasi-stationary distributions (\qsds) of the driving process.
 \par A \qsd is an invariant measure of the driving process conditioned to non-absorption. It is a non-trivial object whose  existence and number are not completely investigated in countable spaces. Besides existence, explicit construction (simulation) of those measures is also a problem, especially if the probability of absorption is very small.
One of the features of the present approach is that it provides  numerical predictions. Using long-time simulations of the \fv particle system, one can obtain immediate valuable  insights about \qsds of $Z_t$.
An excellent overview of the achievements and challenges in the simulation of \qsds is given in \cite{Groisman2012}.

\par The Fleming-Viot approach to the study of \qsds, in the discrete space setting, has been introduced in \cite{Maric2006}, \cite{FerrariM2007}. 
Some questions have been answered for finite space \cite{Asselah2009} and countable space under certain conditions (\cite{FerrariM2007},\cite{Asselah2012}), but there are still many open problems regarding limiting behaviour of the \fv process.  Here we examine perhaps the most paradigmatic and still puzzling case of  a driving process being  a nearest-neighbor random walk on $\N$, with absorption  at origin. This random walk has infinitely many \qsds if there is a drift towards 0, and none otherwise. In the former case it is a one parameter family $\nu_c$, where $0 \leq c < 1-p/q$ and $q,p$ are rates of hopping to the left and to the right respectively.  We will refer to this particle system as FVRW (Fleming-Viot driven by a Random Walk). When $c=0$, the corresponding $\nu_0$ is called the {\it minimal \qsd}. Under this measure, expected time till absorption is minimal, compared to the other \qsds. 

%%%%%%%%%%sad kazem sta je uradjeno a ne radi za rw
\par It has been  recently proved in \cite{Asselah2012} that FVRW is ergodic. Since RW (Random Walk) has infinitely many \qsds, the question is which one is approximated by the mean normalized densities of the FVRW stationary measure. It is believed (\cite{Maric2006}, \cite{Asselah2012}, \cite{GroismanJ2013})  that in the limit, as $N \to \infty$,  is exactly the minimal \qsd:  $\nu_0$.  This property is usually reffered to as a  {\em selection principle}. The analogous result is proven in  the case of subcritical branching process \cite{AsselahFerrari2012}, and some birth and death processes \cite{Villemonais2014} but the methods used there do not apply in the RW case. We use graphical construction of the FVRW and computer simulations to support the above conjecture. 
\par Here we also examine the role of others \qsds,  $\nu_c$ $(c>0)$ in the corresponding FVRW process. We performed simulations drawing starting profiles  from a \qsd that is not the minimal one (independently for each particle). For every combination of parameters $q$ and $c$ we observed significant sojourn time that increases exponentially both with $q$ and $c$.  This feature is typical  for metastability, which leads us to conjecture that each $\nu_c$ $(c>0)$ corresponds to a metastable state of the FVRW, as $N \to \infty$. 

\par The remainder of the paper is organized as follows. Section 2 is devoted to the \qsds of the random walk. In Section 3 we define the \fv process and perform a graphical construction of FVRW. Section 4 contains findings based on simulations. Finally, Section 5 is reserved for a brief discussion.

\section{Quasi-stationary distributions on countable spaces}

Let $Z_t$ be  a pure jump regular Markov process
on a countable $\Lambda\cup\{0\}$ with absorbing state $0$.  For  $x,y  \in \Lambda$  we will denote transition rates matrix  $Q$ ($Q=(q(x,y)): q(x,y)$ is a transition rate from $x$ to $y$) and transition probabilities $P_t(x,y)$. 
Assume also that the exit rates are uniformly bounded above: 
$\bar q:=\sup_{x} \sum_{y\in\{0\}\cup\Lambda\setminus\{x\}} q(x,y)<\infty$, $P_t(x,y)>0$ for
all $x,y\in\Lambda$ and $t>0$ and that the absorption time is almost surely finite for any initial state.  
This type of process is often seen in applications, for example if we consider the spread of an endemic infection, the number of infected individuals of the population could be $Z_t$. Classical Markov theory ensures that there is a unique stationary distribution concentrated at $0$. When the period before the absorption  is extended (but a.s.\ finite), it is interesting to see whether the distribution of the number of infected individuals during this time exhibits a  regular behavior.

Let $\mu$ be a probability on $\Lambda$. The law of the process at time $t$ starting with $\mu$ conditioned to non-absorption until time $t$ is given by
\begin{equation} \label{u7}
  \varphi^\mu_t (x) = \frac{\sum_{y\in\Lambda} \mu(y)
  P_t(y,x)}{1-\sum_{y\in\Lambda} \mu(y) P_t(y,0)}\, .
\end{equation}

A \emph{quasi stationary distribution} ({\qsd}) is a probability measure $\nu$
on $\Lambda$ satisfying $\varphi^\nu_t = \nu$, that is: an invariant measure for the conditioned process. 
 A {\qsd} is a left eigenvector $\nu$ for the restriction of the matrix $Q$ to $\Lambda$ with eigenvalue: 
$-\sum_{y\in\Lambda} \nu(y) q(y,0)$. That is, $\nu$ must satisfy the system
\begin{eqnarray}\label{eq:quasi2}   
  \sum_{y \in \Lambda}\nu(y)\,[q(y,x)+q(y,0)\nu(x)]=0 , ~~\forall x \in \Lambda.   
\end{eqnarray} 
(Recall $q(x,x) = -\sum_{y\in\Lambda\cup\{0\}\setminus\{x\}} q(x,y)$.)
\\

So, finding a \qsd involves solving a system of non-linear equations which is a difficult task, in general. However, in the case of the random walk that we consider here, we get a system of difference equations that is solvable using standard methods.

\subsection{ \qsds for random walk on $\mathbb{N}$ with absorption at 0} 
Consider a continuous-time random walk on  $\N$ with an absorbing barrier at 0:
$q(x,x-1)=q$, $q(x,x+1)=p$,  and $q(0,0)=0$. We will additionally assume that there is a drift towards 0, namely that $q>p$, since otherwise there  is no  a \qsd \cite{Cavender1978}.
\par A \qsd for this process satisfies the equation (\ref{eq:quasi2}):
\begin{eqnarray}\label{sistemqp} 
\nu(x)= \frac{\nu(x-1)p + \nu(x+1)q}{p+q-\nu(1)q},  ~~x \geq  2. 
\end{eqnarray} 
%Let $\lambda(\nu)=(1-\nu(1)q)$ . 
Then we have  homogeneous difference equations of the second order: 
\begin{eqnarray*} 
\nu(2)= \frac{1}{q}(p+q-\nu(1)q) \nu(1) 
\end{eqnarray*} 
and
\begin{eqnarray*} 
\nu(x)- \frac{(p+q-\nu(1)q)}{q}\nu(x-1) + \frac{p}{q}\nu(x-2)=0 ~~  \mbox{for}~ x \geq 3. 
\end{eqnarray*} 
 Define $c= [(\nu(1)-p/q-1)^2 -4p/q] ^{1/2}$. Then the characteristic equation 
 \begin{eqnarray*}
 z^2-\frac{(p+q-\nu(1)q)}{q} z + \frac{p}{q}=0
 \end{eqnarray*}
 has the following solutions
\begin{eqnarray*} 
z_{1,2}=\frac{(p+q-\nu(1)q)/q \pm  \sqrt{(p+q-\nu(1)q)^2/ q^2 - 4p/q}}{2}
=\frac{c \pm  \sqrt{c^2 + 4p/q}}{2} 
\end{eqnarray*} 
and  the equation has real solutions for $(p+q-\nu(1)q) \geq 2\sqrt{pq}$ or $\nu(1) \leq (\sqrt{p/q}-1)^2 $. Considering that $\nu(1)$ has also to be strictly positive, the last condition is equivalent to $ 0 \leq c <  |1- p/q|$. Recall that $p<q$, so the last condition is actually 
\begin{equation}\label{rangec}
0 \leq c < 1-p/q.
\end{equation}
 
 The minimal value $c=0$ would correspond to the minimal \qsd. Then, 
$\nu_{0}(1)=(\sqrt{p/q}-1)^2$ and there is only one root of the above equation, $z=(p+q-\nu_{0}(1)q)/2q = \sqrt{\frac{p}{q}}  $. In that case the solution has the form
$\nu_{0}(n)= z^{n}(a + bn)$. The constants, found from $\nu_{0}(1)$ and $\nu_{0}(2)$, are

$a=0, b=(p+q)/\sqrt{pq} - 2$ and the general solution is given by
\begin{equation} \nonumber
\nu_{0}(n)= \Big(\frac{p+q}{\sqrt{pq}} - 2\Big)n\Big(\sqrt{\frac{p}{q}} \Big)^n= \nu_0(1) n \Big(\frac{p/q+1 - \nu_0(1)}{2}\Big)^{n-1} .
\end{equation} 

\begin{figure}[H] 
\centering 
\includegraphics[width=4in]{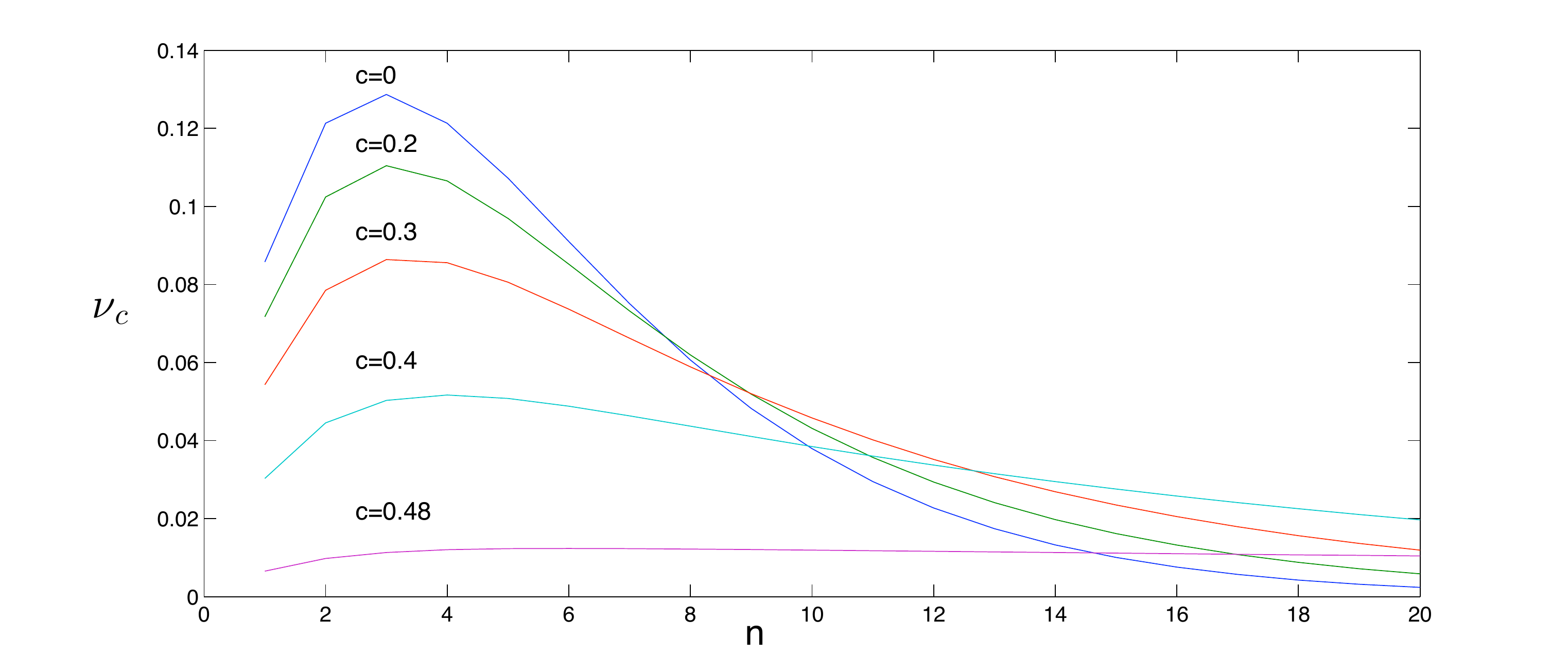} 
 \caption{Quasi-stationary distributions  for RW on $\N$ with $q=2/3, p=1/3$. They are parametrized by $c: 0 \leq c < 0.5$ } 
 \label{infinitas} 
\end{figure}

For $c>0$ we will obtain another solution for the system (\ref{sistemqp}) which would also be a \qsd.  So there is an entire family of \qsds parametrized by $c$, and they have the following form
\begin{equation}\label{qsdc}
\nu_c(n)= \frac{\nu_c(1)}{c} \Big[ \Big(\frac{p/q+1 - \nu_c(1)+c}{2}\Big)^n - \Big(\frac{p/q+1-\nu_c(1)-c}{2}\Big)^n \Big].
\end{equation}
Observe that as $c$ increases, $\nu_c(1)$ gets smaller which can also be seen in Figure \ref{infinitas}
%%%%%%%%%%%% sliku zameniti %%%%%%%%%%%%%%%%%%%%%

\section{Fleming-Viot particle system driven by a Random Walk (FVRW)}
\emph{The Fleming-Viot process} ({\fv}).  
Consider a system of $N$ particles ($N \geq 2$) evolving on a countable space $\Lambda$.  The particles move
independently, each of them governed by the transition rates $Q$ until absorption. Since there cannot be two simultaneous jumps, at most one particle
is absorbed at any given time.  When a particle is absorbed to 0, it goes instantaneously to a site in $\Lambda$ chosen with the empirical distribution
of the particles remaining in $\Lambda$.  In other words, it chooses one of the other particles uniformly at random  and jumps to its position.  Between absorption times
the particles move independently, governed by $Q$. 

This process has been
initially studied 
in a Brownian motion setting \cite{Burdzy2000}.  Countable space has been treated firstly in \cite{Maric2006}, \cite{FerrariM2007} and later in \cite{AsselahFerrari2012}, \cite{Asselah2009}, \cite{Asselah2012}. The original process introduced by Fleming and Viot \cite{FlemingV1979} is a model for a population with constant number of individuals which also encodes the positions of particles. 

The generator of the \fv process acts on functions
$f:\Lambda^{(1,\dots,N)}\to\R$ as follows
\begin{equation} \label{a5}
  \cL^N f(\xi) = \sum_{i=1}^N \sum_{y\in\Lambda\setminus\{\xi(i)\}}
  \Bigl[q(\xi(i),y) + q(\xi(i),0)\, \frac{\eta(\xi,y)}{N-1}\Bigr] (f(\xi^{i,y}) - f(\xi)),
\end{equation}
where $\xi^{i,y}(j) = y$ for $j=i$
and $\xi^{i,y}(j) = \xi(j)$ otherwise and
\begin{equation} \nonumber
  \eta(\xi,y) := \sum_{i=1}^N \one\{\xi(i)=y\}. 
\end{equation}
Namely, $ \eta(\xi,y)$ is the number of particles at site $y$, in the configuration $\xi$.
We call $\xi_t$ the process in $\Lambda^{(1,\dots,N)}$ with generator \reff{a5}
and $\eta_t = \eta(\xi_t,\cdot)$ the corresponding unlabeled process on
$\{0,1,\dots\}^\Lambda$; $\eta_t(x)$ counts the number of $\xi$ particles in
state $x$ at time $t$. 

For $\mu$ a measure on $\Lambda$, we denote by
$\xi^{N,\mu}_t$ the process starting with independent identically
$\mu$-distributed random variables $(\xi^{N,\mu}_0(i),\,i=1,\dots,N)$; the
corresponding variables $\eta^{N,\mu}_0(x)$ follow a multinomial law with parameters
$N$ and $(\mu(x),\,x\in\Lambda)$. 

%Theorem 1.2 in \cite{FerrariM2007} says that the profile of the {\fv} process at time $t$
%converges, as $N \to \infty$, to the conditioned evolution of the chain $Z_t$,  $\varphi_t $. 

%

\subsection{Construction of FVRW process}

Graphical representation in interacting particle systems has been used extensively since the pioneering work of Harris \cite{Harris1978}. The main idea is to construct the process explicitly in terms of independent collections of Poisson processes \cite{Liggett2011}. Along these lines, here we perform a graphical representation/construction of the FVRW process $\xi_t^N$. 
Events in Poisson processes correspond to clocks according to which the particles jump, and we will call them {\em internal times}. When a clock goes off, random variables called {\em marks} are used to determine next position of the particle. 
For each $i=1,\dots,N$, we define independent stationary marked Poisson processes on $\R$:
\begin{itemize}  

\item {\em Internal times:} Poisson process with rate $q+p$: $(b^{i}_n)_{n \in\Z}$, with marks
  $((B^{i}_n(x),\,x\in\Lambda),\,n \in\Z)$,
  $((C^{i}_n),\,n \in\Z)$. %and $((U^{i}_n),\,n \in\Z)$.
\end{itemize}

The marks are independent of the Poisson processes and mutually independent. Their marginal laws are given below:
\begin{itemize}

\item $\P(B^{i}_n(x)= x+1)= \frac{p}{p+q}$,\\
$\P(B^{i}_n(x)= x-1)=  \frac{q}{p+q} $, $ x \in \N$.
\item $\P(C^{i}_n=j)= \displaystyle{\frac{1}{N-1}}$, $j\neq i$.
%$\item $U^i_n \sim Uniform(0,1)$
\end{itemize}

Denote by $(\Omega,\mathcal{F}, \P)$ the probability space on which the marked Poisson processes have been constructed.  Discard the null event corresponding to two
simultaneous events at any given time.

We construct the process in an arbitrary time interval $[s,t]$. Given the mark
configuration $\omega \in \Omega$ we construct $\xi_{[s,t]}^{N,\xi}(=\xi^{N,\xi}_{[s,t],\omega})$ in the time interval $[s,t]$ as a
function of the Poisson times, and their respective marks, and the initial configuration $\xi$ at time $s$.

\paragraph{Construction of $\xi_{[s,t]}^{N,\xi}=\xi_{[s,t],\omega}^{N,\xi}$ }\
 
Since for each particle $i$ there is a Poisson process, the number of events in the interval $[s,t]$ is
Poisson with mean $N(p+q)$. So the events can be ordered from the earliest
to the latest. If at time $s$ the initial configuration is $\xi$, then, we proceed event by event
following the order as follows (between Poisson events the configuration does not change):\\
If at the internal time $b^i_n-$ the state of particle $i$ is $x$, and $x\neq 1$ then at time
$b^i_n$ particle $i$ jumps to state $B^i_n(x)$ regardless of the position of the
other particles. If $x = 1$ and $B^i_n(1)=0$, particle $i$ jumps to the site where  particle $C^i_n$ is; if
$B^i_n(1)=2$, then the state of particle $i$ becomes 2. The configuration obtained after using all events is $\xi_{[s,t]}^{N,\xi}$.

The above graphical construction is algorithmized in the Algorithm FVRW:

\subsection{Algorithm FVRW}
\begin{itemize}
\item[\bf Step 1] T=0; Sample $\xi_0(j) \sim \mu$, $j=1,...,N$;  \\
 Set $\eta_0(x)=\sum_{i=1}^N \one\{\xi_0(i)=x\}, x=1,2,... $\\
\item[\bf Step 2] Sample $t \sim Exponential(q+p)$.\\
                             Choose particle $i$ uniformly at random from $\{1,...,N\}$.\\ 
                             Number of particles at the site $\xi_T(i)$ is decreased by 1: 
                             $\eta_{T+t}(\xi_T(i))=\eta_T(\xi_T(i))-1$.\\
\item[\bf Step 3] 
Sample $U \sim Uniform(0,1)$. \\
-If $ U<q/(q+p)$
\begin{itemize}
\item if $\xi_T(i)=1$  then %"choose one of the remaining particles and jump to its position"\\
choose particle $j$ uniformly at random from $\{1,...,i-1,i+1,...,N\}$.\\
Particle $i$ jumps to the position of particle $j$: $\xi_{T+t}(i)=\xi_T(j)$
and number of particles at the site $\xi_T(j)$ increases by 1: $\eta_{T+t}(\xi_T(j))=\eta_T(\xi_T(j))+1.$
 \item if $\xi_T(i) \neq 1$  then particle $i$ jumps one position to the left and the number of particles at the new site gets updated: $\xi_{T+t}(i) = \xi_T(i)-1$; $\eta_{T+t}(\xi_T(i)-1)= \eta_{T}(\xi_T(i)-1)+1$.
\end{itemize}
\vskip3mm
 -If $U>q/(q+p)$ (i.e. with probability p/p+q)\\
 then particle $i$ jumps one position to the right and the number of particles at the new site gets updated:
$\xi_{T+t}(i)=\xi_T(i)+1$; $\eta_{T+t}(\xi_T(i)+1)=\eta_T(\xi_T(i)+1)+1$.\\

\item[\bf Step 4] $T \leftarrow T+t$.  If $T< \tau$ go to Step 2; otherwise STOP.

The output of the algorithm is $\xi^{N,\mu}_{[0,\tau]}$.

\end{itemize}

\section{Findings and Conjectures}

In this section we present the findings based on the simulations performed using MATLAB. Our focus is on getting an insight into qualitative rather than quantitative properties of the FVRW process.
Let us define the mean normalized density as 
\begin{equation} \nonumber
\rho^{N,\mu}_t(k)= \E \frac{\eta_t^{N,\mu}(k)}{N}, ~~ k \in \N
\end{equation}
 where the initial position of all particles is chosen independently with distribution $\mu$.
 We will use further the notation $\rhohatt$ for the estimated density  at time $t$ using a Monte Carlo method. It is obtained as an average, over 50 independent realizations of $\eta_t$, generated by the Algorithm FVRW. 
 
\subsection{Selection principle} 

It was conjectured in \cite{Maric2006}  that FVRW with $(q>p)$ is ergodic, a result which has been recently proved in \cite{Asselah2012}.  
Let $\rho^N=\rho^N_\infty$ be the density of the \fv process in equilibrium (note that initial configuration does not play a role here so we omit it from the notation).  It has been also conjectured in \cite{Maric2006} that as $N$ goes to infinity this empirical equilibrium density approaches the {\it minimal \qsd}.
\begin{conj} [ Mari\'c \cite{Maric2006}] For the \fv driven by RW on $\N$, with $q>p$
\begin{eqnarray*}
\rho^N \to \nu_{0}, ~~N \to \infty
\end{eqnarray*}
\end{conj}

\par Heuristic arguments are based on the following two facts: 1. $\rho^{N,\mu}_t$ converges to $\varphi_t$ (defined in \reff{u7}) as $N \to \infty$  (Theorem 1.2 in ~\cite{FerrariM2007}).
2.  $\varphi_t$ converges to the minimal \qsd $\nu_0$, as $t \to \infty$ \cite{Ferrari1995}.
 
\par Analogous result was proven for the \fv  driven by a subcritical Galton-Watson \cite{AsselahFerrari2012} and very recently, for those driven by some birth and death processes (not including a random walk)\cite{Villemonais2014} .
In what follows we are going to provide simulation evidence in support of the above Conjecture. 

\par Let $\rhohat$ be the estimated limiting density profile, obtained as an average, over 50 independent realizations of $\eta_t$, obtained using the Algorithm FVRW.  Initially all the particles are positioned at 5 and $t$ is chosen large enough that  we may say the equilibrium distribution has been reached.  The initial position is chosen to be 5 without any special reason, since the process is ergodic, initial configuration does not affect long-time behavior.

\begin{figure}[H] 
\centering 
\includegraphics[width=4in]{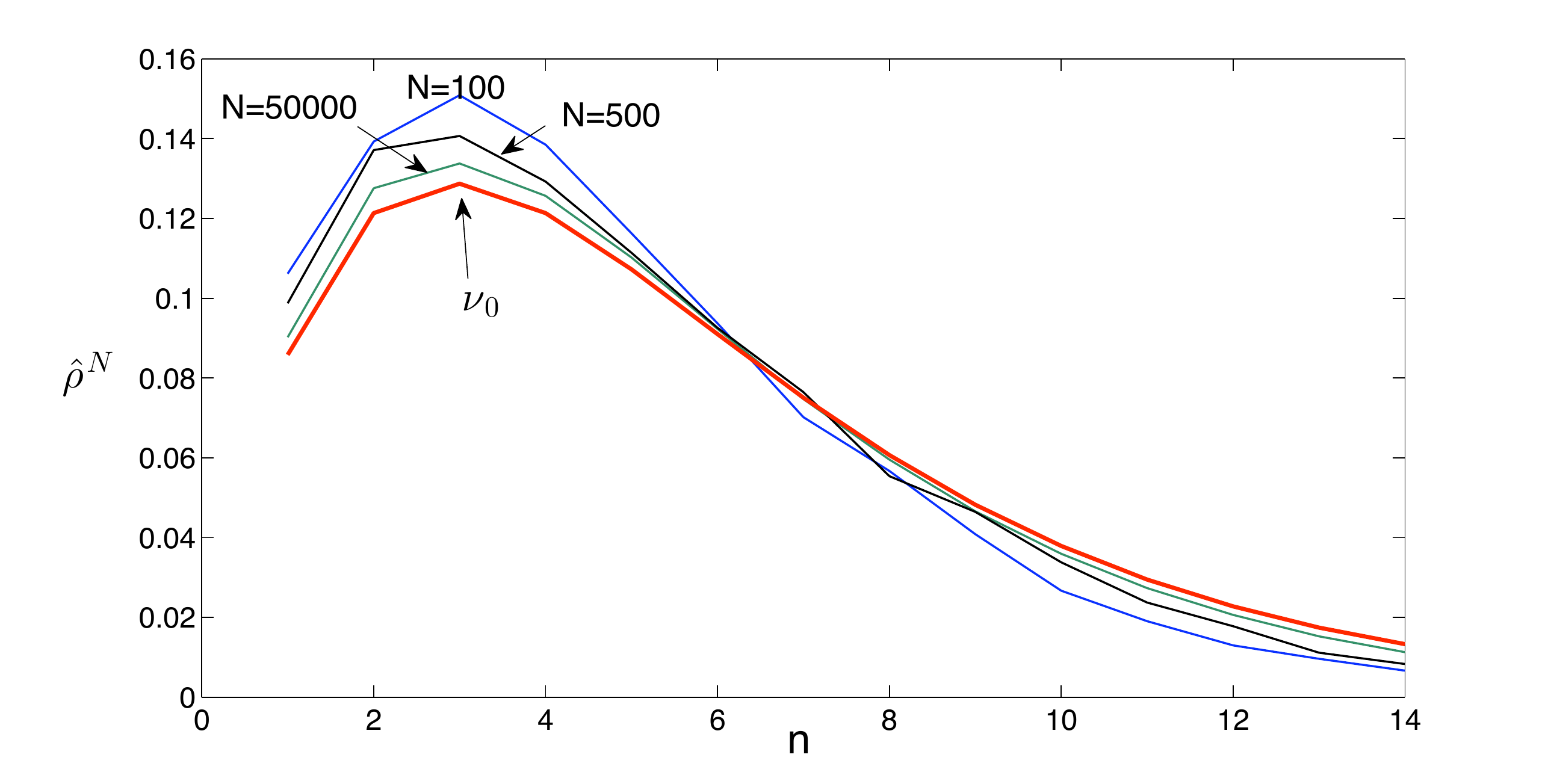}
 \caption{ Red curve is the minimal \qsd $\nu_0$. Three other curves, obtained from simulations, present $\hat \rho^N$, for $N=100, 500, 50000$. } 
  \label{selpric1} 
\end{figure}

 Figure \ref{selpric1}  compares  minimal \qsd with the limiting FVRW densities for $N= 100, 500, 10000$.  Note how, with larger $N$, the approximation by $\nu_0$ becomes better.

\par Let us define the {\it L-truncated total variation distance} between two probability measure $\mu$ and $\nu$ as  $d_{TV}^L(\mu,\nu) = 1/2 \sum _{i=1}^{L} |\mu(i)-\nu(i)|$.  The truncation is necessary since we are looking at the finite window of  infinite-volume measures. Then we obtain $\rhohat$ for different values of $N$ in the range $(500,1000,..., 50000)$ and show the distance $d_{TV}^L(\rhohat,\nu_0)$, for $L=150$.  Note that, when truncated at  $L=150$, measure $\nu_0$, is still very well approximated.  For example, for $q=2/3, p=1/3$, $\nu_c(100) \sim 10^{-14}$, and $\nu_c$ is decreasing in $n$ for $n>3$. Consequently, $d_{TV}^{150}(\rhohat,\nu_0)$ is very close to $d_{TV}(\rhohat,\nu_0)$, (non-truncated) total variation distance.

\begin{figure}[H] 
\centering 
\includegraphics[width=4in]{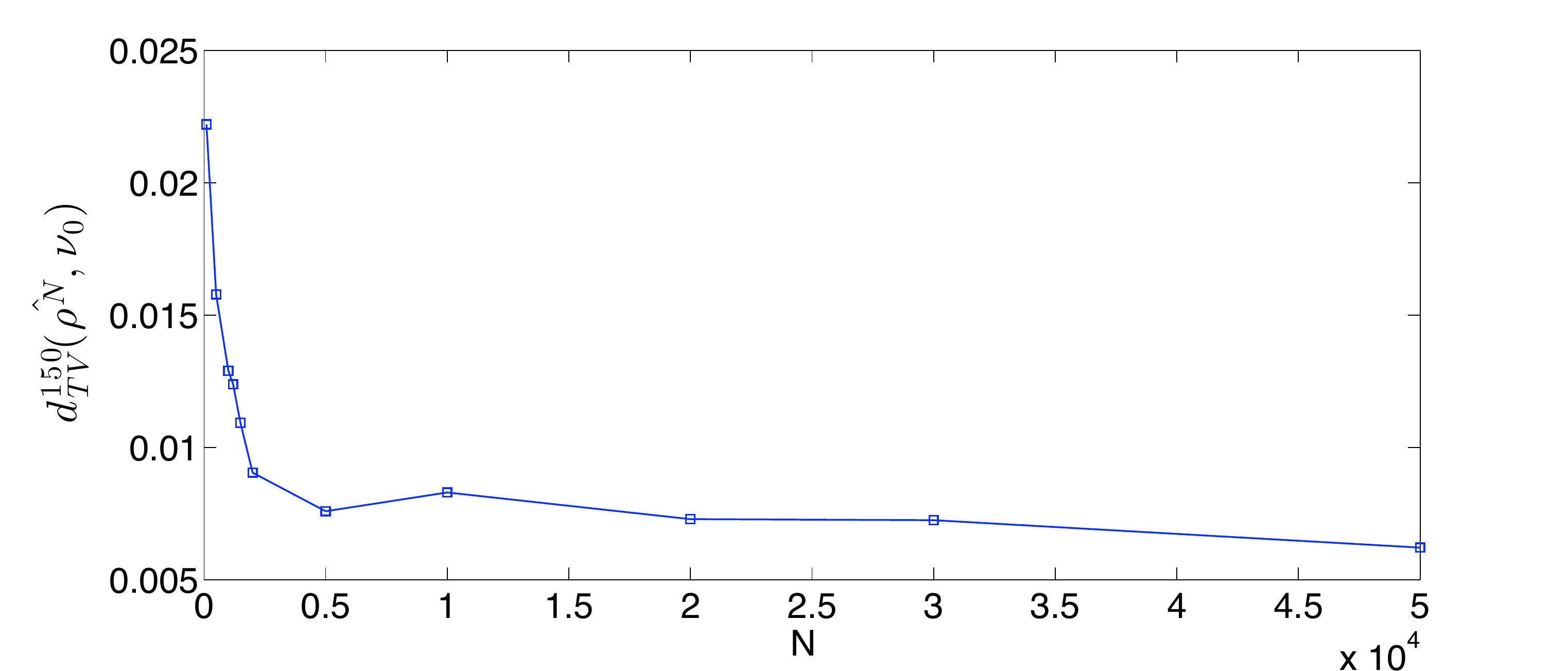}
 \caption{150-truncated total variation distance between $\rhohat$ and $\nu_0$ is approaching $0$ as $N$ gets larger. Here $q=2/3, p=1/3$.} 
  \label{selpric2} 
\end{figure}

In Figure \ref{selpric2} are presented the obtained values for $q=2/3,p=1/3$. It clearly supports the conjecture that $d_{TV}(\rhohat,\nu_0) \to 0$ as $N \to \infty$.

\subsection{Metastability}

Since the minimal \qsd is the limiting density of the FVRW process it is natural to ask whether there is a special meaning of other \qsds for this particle system. What happens if initially each particle's position is chosen according to another \qsd (not the minimal one)?

\par Suppose then that initially each particle is positioned, independently of others, according to $\nu_c$ ($c>0$), where $\nu_c$ given by \reff{qsdc}, is a \qsd for the random walk on $\N$.  Although $\nu_c$ is an infinite-volume measure, the number of particles $N$ is finite, and the entire system is therefore finite. The position of  right-most particle changes, it can be very far from the origin, but at any given time it is finite. 

Now, let $\hat\rho^{N,c}_t$ be the estimated mean density (via Monte Carlo) of the FVRW at time $t$. Let also  $\hat\rho^{N,c}$ be the Monte Carlo estimated mean stationary (steady) state, which exists by the arguments mentioned at the beginning of this section.
In order to monitor velocity of convergence  we plot 
$d_{TV}({\hat\rho_t^{N,c}}, {\hat\rho^{N,c}})$ 
as a function of $t$.  Note that both ${\hat\rho_t^{N,c}}$ and $\hat\rho^{N,c}$ are finite-volume probability measures and there is no need here to use truncated total variation distance.

\begin{figure}[H] 
\centering 
\includegraphics[width=4in]{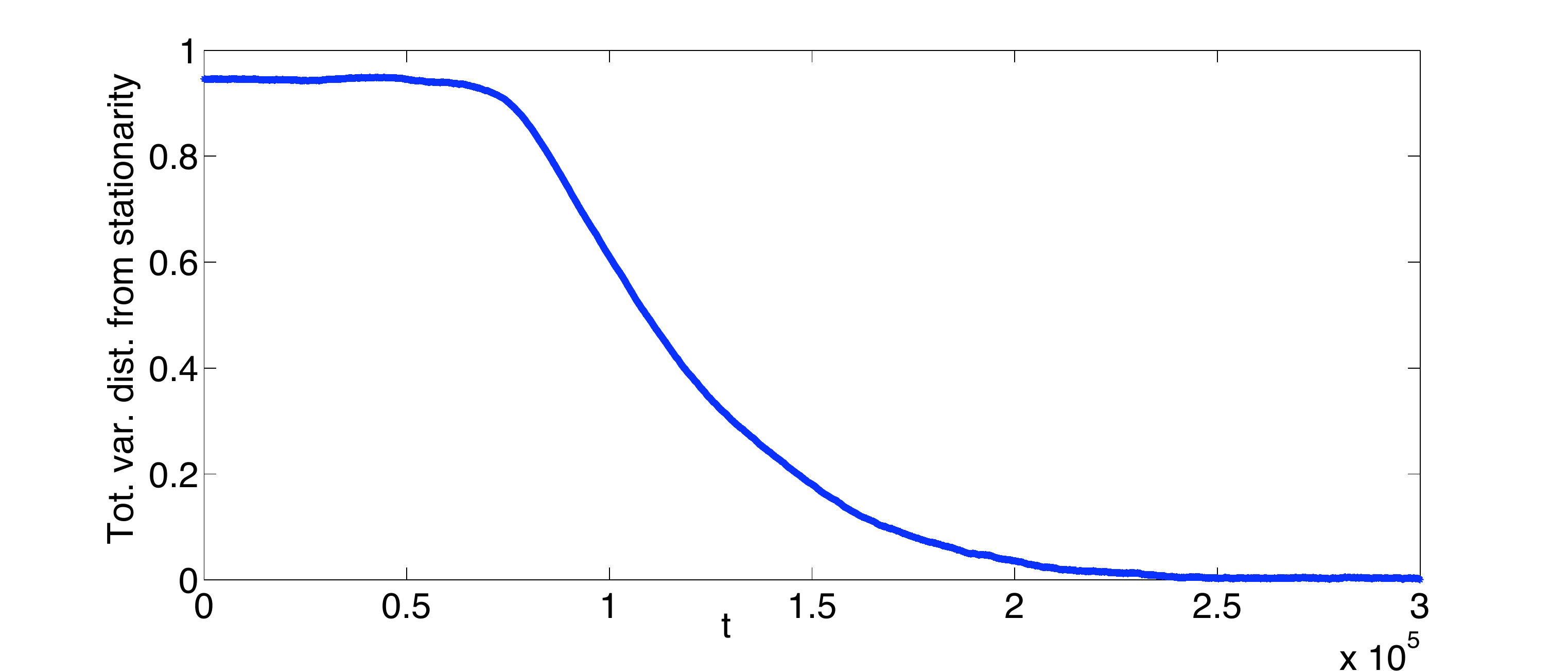} 
 \caption{Total variation distance from the steady state $d_{TV}({\hat\rho_t^{N,c}}, {\hat\rho^{N,c}})$ as a function of $t$.  Here q=4/5; c=0.6; N=50000. } 
 \label{onepl} 
\end{figure}

Figure \ref{onepl} displays a typical result, where $N=50000, q=4/5, c=0.6$.

{\bf Remark}: In all figures in this section where $x$-axis is labeled with $t$,  the time is scaled. Real simulation time is 100 times larger.

\par We have performed numerous simulations changing all three parameters ($q, c, N$). In each case that we analyzed, a {\it plateau} was present, meaning that the process stays a ``long" time in the initial distribution. Then, it starts ``rapidly" to approach the equilibrium. This type of behavior  is typical for metastability (\cite{Liggett1999}, \cite{Rakos2003}), as is the requirement that the plateau length be approximately an exponential function of a certain parameter of the system. 

\par For any $q > 1/2$, define the plateau length as 
\[ 
PL_{c,q}^N(\epsilon) = \min\{t: d_{TV}({\hat\rho_t^{N,c}}, {\hat\rho^{N,c}_0})> \epsilon\}
\]
where $\epsilon$ is positive but relatively small.  Typically, we take $\epsilon = 0.005$. In what follows, we compare $PL_{c,q}^N$ for different values of $N, c, q$ respectively. Each time, the other two parameters stay fixed. 

\noindent $\bullet$ {\bf Plateau length as a function of  $N$}
\begin{figure}[H] 
\centering 
\includegraphics[width=5in]{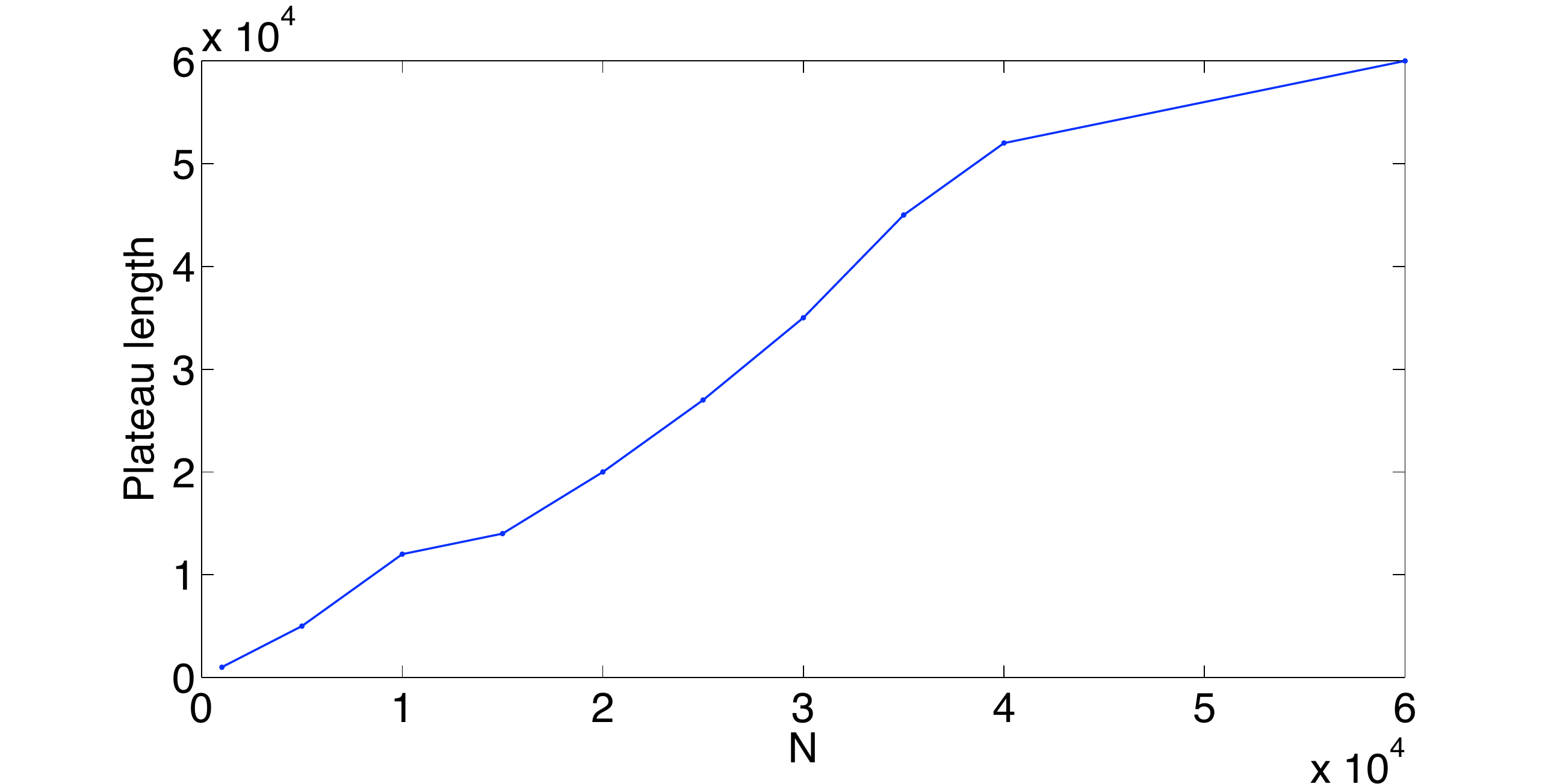} 
 \caption{Plateau length $PL_{c,q}^N(0.005)$ as a function of number of particles N. Here  $q=0.8, c=0.6$} 
 \label{diffN} 
\end{figure} 
As we can see on the Figure \ref{diffN}, growth in  $PL_{c,q}^N$ with $N$ is observed but it is very slow, almost linear.  It is somehow not surprising that with increase in $N$ there is no dramatic change in the behavior of the process. 
Having in mind arguments around the Selection principle we would expect that a \qsd of the random walk  $\nu_c$ has a special meaning (if any) for the FVRW process in the limit as  $N \to \infty$. 

\noindent $\bullet$ {\bf Plateau length as a function of $c$}

\begin{figure}[H] 
\centering 
\includegraphics[width=5in]{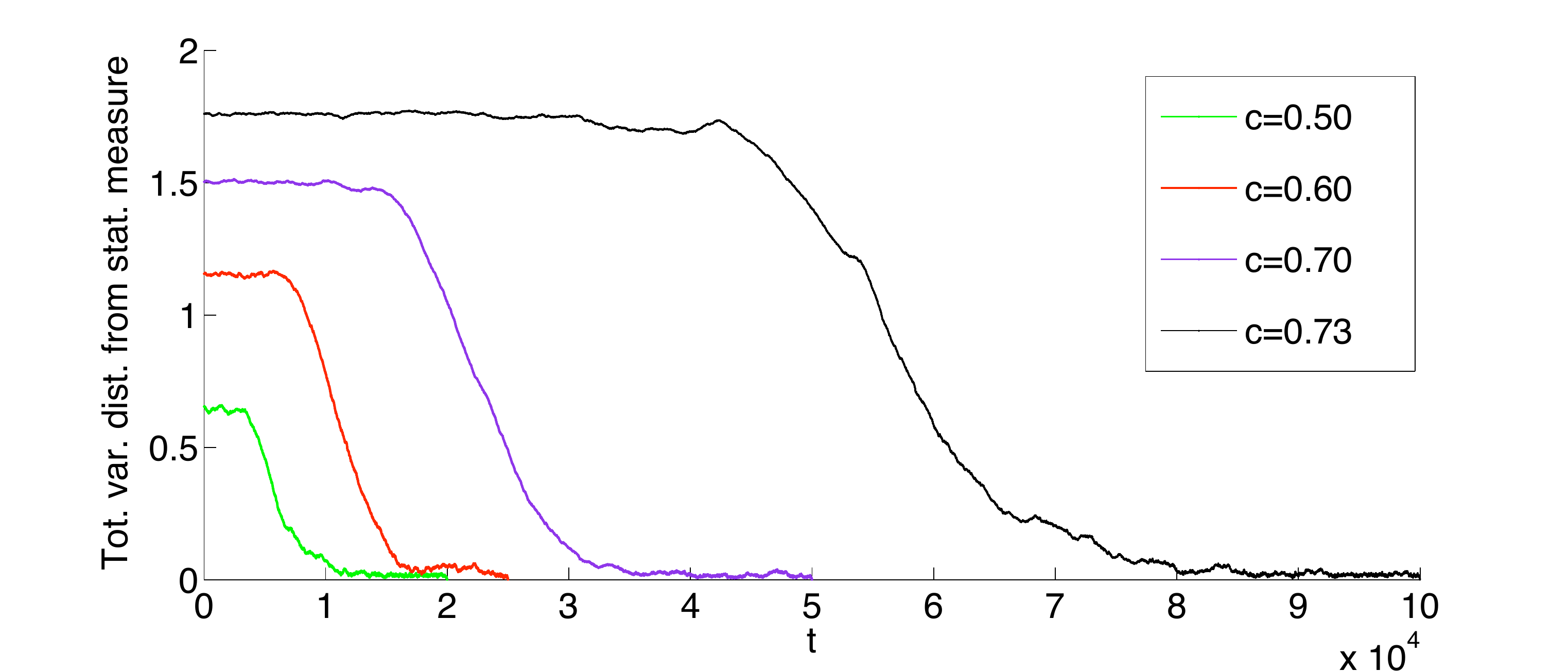}
 \caption{$d_{TV}({\hat\rho_t^{N,c}}, {\hat\rho^{N,c}})$ for $c=0.5, 0.6, 0.7, 0.73$. Here  q=0.8, p=0.2; N=5000.} 
  \label{plat} 
\end{figure}

\begin{figure}[H] 
\centering 
\includegraphics[width=5in]{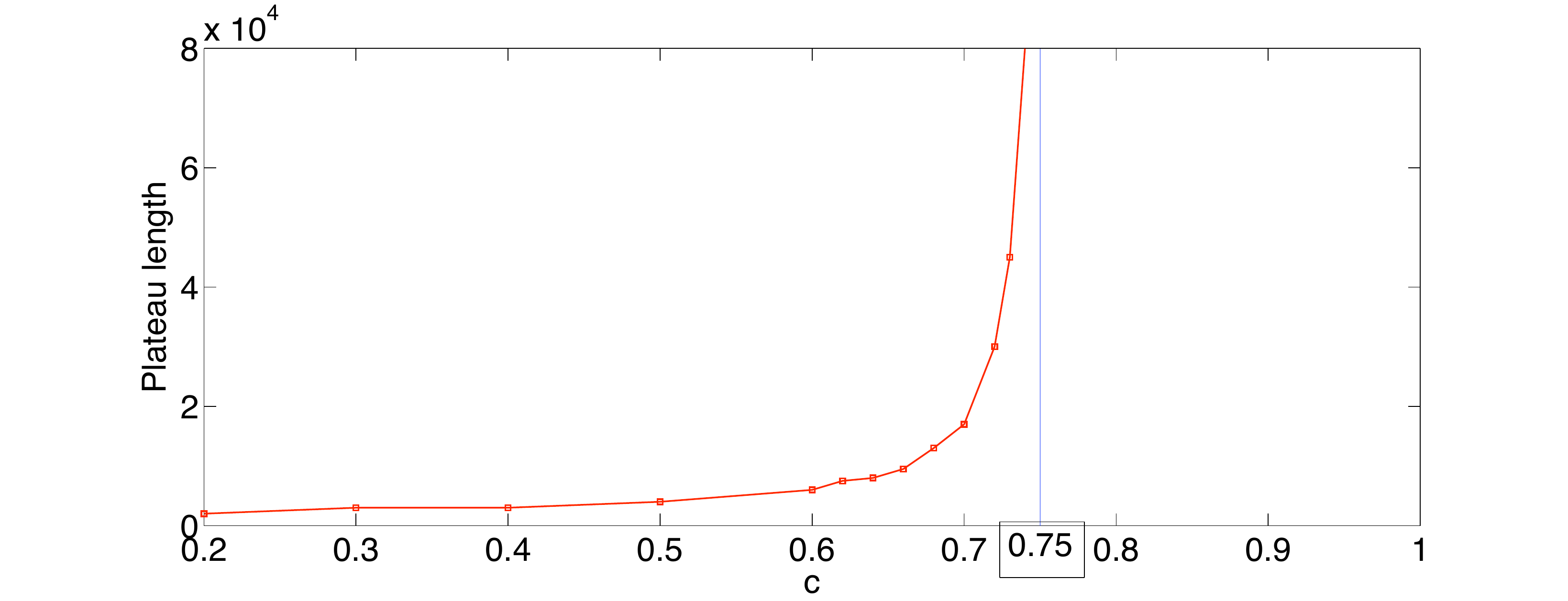} 
 \caption{Plateau length $PL_{c,q}^N(0.005)$  for different values of c =0.2,...,0.73. Here  q=0.8, p=0.2; N=5000. For q=0.8, p=0.2  the condition \reff{rangec} gives a range  for $c$: $0\leq c < 0.75$.
 } 
 \label{difc} 
\end{figure}

As mentioned in the Section 2, for every choice of $q$ and $p$ ($q>p$), there is an entire family of \qsds, parametrized by $c$ where $0 \leq c < 1-p/q$. 
 In Figure \ref{plat} is displayed  $d_{TV}({\hat\rho_t^{N,c}}, {\hat\rho^{N,c}})$ for different values of c. Obviously plateau lengths increase with $c$.  
A more detailed comparison, showing $PL_{c,q}^N$ as a function of $c$, is given in Figure \ref{difc}.
From Figure \ref{difc} it is clear that the length of the plateau increases significantly with $c$ and appears to diverge as $ c \to 0.75$. For the particular choice of parameters $q=0.8, p=0.2$,  value 0.75 corresponds to the theoretical bound for $c$ as given in condition \reff{rangec}.
 \par
\begin{figure}[H] 
\centering 
\includegraphics[width=5in]{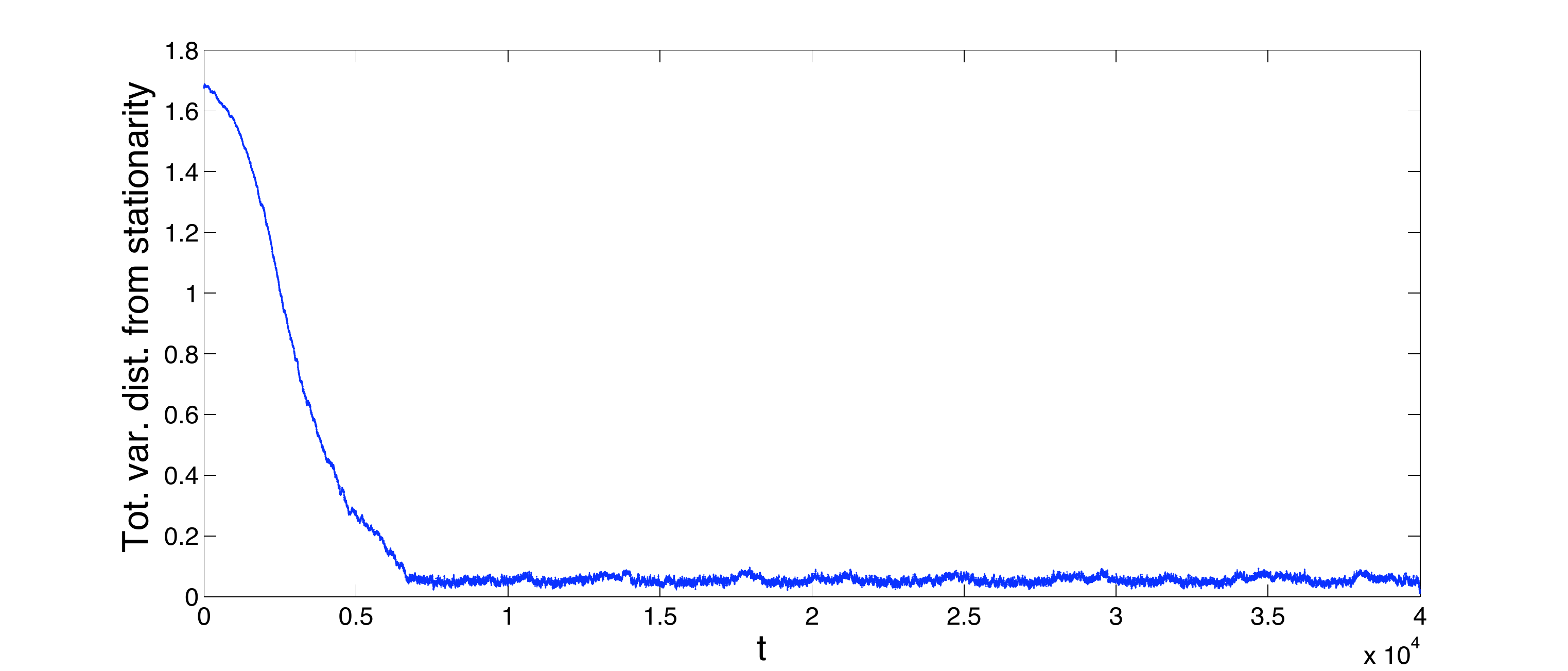}
 \caption{$d_{TV}({\hat\rho_t^{N,\mu}}, {\hat\rho^{N,\mu}})$ as a function of $t$, where $\mu$ is a uniform distribution on [1,100]; Here N=1000; q=2/3; p=1/3. No plateau.} 
  \label{flat} 
\end{figure}

 At the same time, as $c$ increases, $\nu_c$ becomes more ``flat"  with heavier tails (see Figure \ref{infinitas}). One could think then that the main feature affecting plateau lengths is ``flatness" of the initial profile. For that reason we performed simulations using as initial distribution the uniform distribution over large interval $[0,100]$. This initial profile, of course, does not correspond to any $\nu_c$.  It may be clearly seen in Figure \ref{flat}, that in this case convergence is very fast and no plateau is observed whatsoever.
\noindent $\bullet$ {\bf Plateau length as a function of $q$}
\par
\begin{figure}[H] 
\centering 
\includegraphics[width=5in]{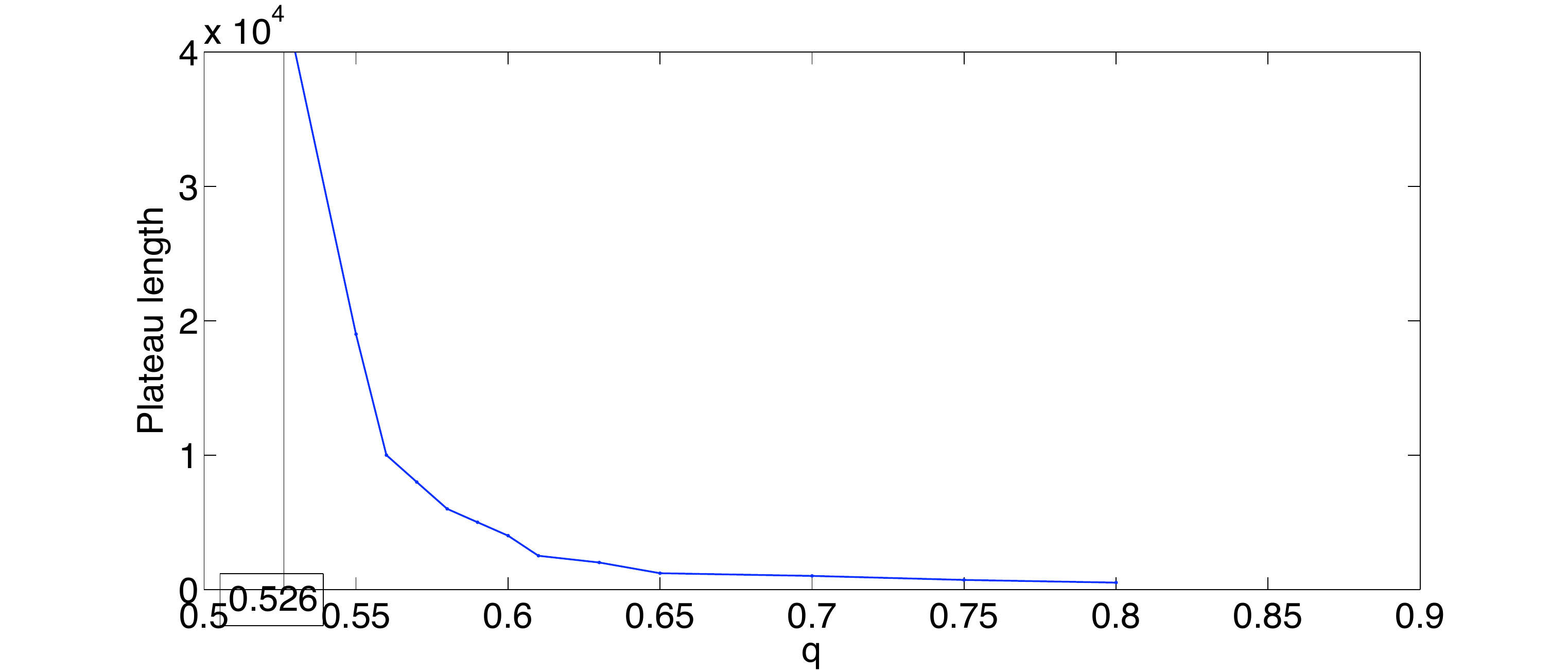}
 \caption{Plateau length $PL_{c,q}^N(0.005)$  for different values of  q=0.8,...,0.53; Here c=0.1, N=1000. Value $0.5263$ corresponds to the theoretical bound for $q$ (given $c$). } 
  \label{difq} 
\end{figure}

\begin{figure}[H] 
\centering 
\includegraphics[width=5in]{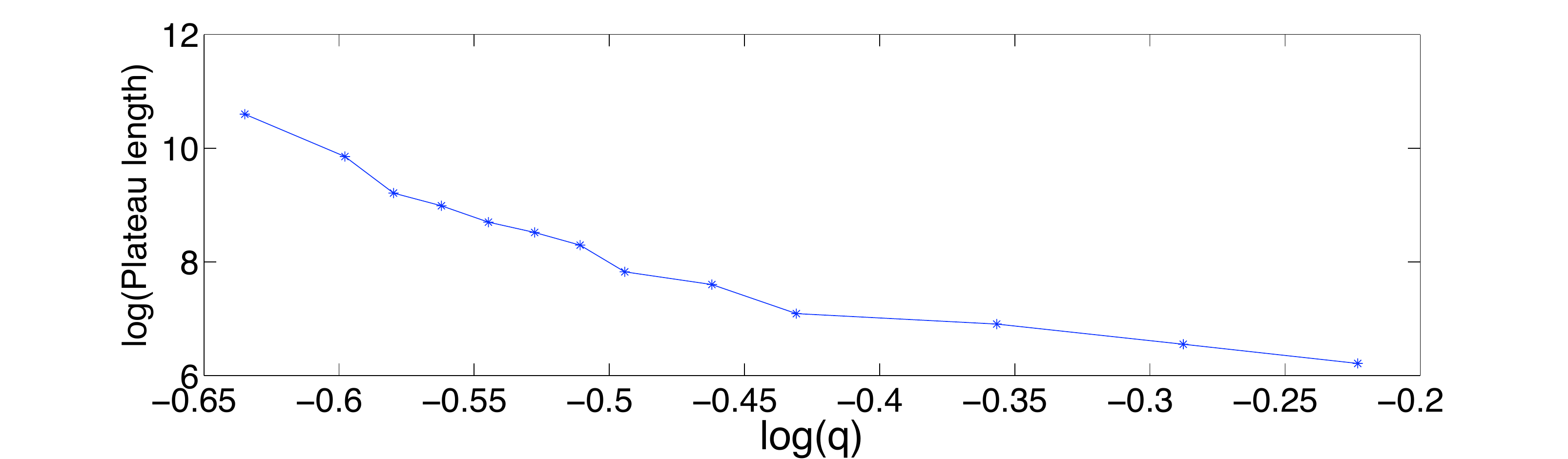}
 \caption{Log-log plot: q =0.8,...,0.5263;  c=0.1, N=1000; } 
  \label{loglogq} 
\end{figure}

 Lastly, we compare sojourn times (plateau lengths) as a function of the rate $q$ (actually the ratio $q/p$, but  in our simulations it is taken $p=1-q$). We approximate plateau lengths for different values of $q$  and fixed value of $c=0.1$. From condition \reff{rangec} we know that the minimum value of $q$ that allows $\nu_{0.1}$ is $q=0.5263$. The results are shown graphically in Figure \ref{difq}. The log-log plot is shown in Figure \ref{loglogq}.

In Figures \ref{difq} and \ref{loglogq}  one can see  that $PL_{c,q}^N$ grows  exponentially as $q$ decreases towards $0.5$.   Recall that in the case $q \leq 0.5$, a random walk does not have ANY \qsd.  The exponential growth of $PL_{c,q}^N$ in $c$ and $q$ is a strong evidence of metastability.  We conjecture that FVRW indeed has uncountably many metastable states. As $N \to \infty$ each metastable state corresponds to  a \qsd $\nu_c$.

\section{Discussion}
In this paper we have studied a Fleming-Viot particle system driven by a random walk on $\N$. The simulations are performed based on the Algorithm FVRW that arose from a graphical construction of the process. It would be very interesting to obtain better analytical results in any  of the directions pursued here. Our findings strongly suggest that mean normalized densities of the FVRW process converge to the minimal quasi-stationary distribution of the random walk, $\nu_0$. This property is often referred to as {\it selection principle}.
\par Furthermore, we presented evidence that FVRW exhibits metastability phenomena.  Moreover, it has uncountable many metastable states. In the case of infinitely many particles, any  \qsd of the random walk: $\nu_c$, $ c>0$, corresponds to a metastable state of the FVRW particle system. These results give a new, {\it physical interpretation} of $\nu_c$, otherwise lacking in the literature. 
These theoretical distributions can now be seen in relation to dynamics of particle systems providing clear physical understanding and making them more applicable for problems in statistical physics.

\vskip5mm
{\bf Acknowledgements}
We are thankful to Gunter Sch\"{u}tz for inspiring discussions. Support from the 
National Science Foundation (grant DMS - 1007823) is gratefully acknowledged.

\bibliographystyle{plain}

\bibliography{My-Collection}   % name your BibTeX data base\end{document}

\end{document}